\documentclass[%
 reprint, 
 amsmath,amssymb,
 aps, prd, longbibliography
]{revtex4-1}

\usepackage{graphicx}% Include figure files
\usepackage{dcolumn}% Align table columns on decimal point
\usepackage{bm}% bold math
\usepackage{tabularx}
\usepackage{natbib}
\usepackage{hyperref}
\usepackage{xcolor}

\def\DeltanuRF{\Delta\nu_{\lower1.5pt\hbox{$\scriptstyle RF$}}}
\def\Eobs{\vec{E}_{\lower0.5pt\hbox{\scriptsize obs}}}
\def\kB{k_ {\lower1.5pt\hbox{$\scriptstyle B$}}}
\def\Psig{P_{\hbox{\scriptsize sig}}} 
\def\Psys{P_{\hbox{\scriptsize sys}}} 
\def\rhoDM{\rho_{\lower1.5pt\hbox{\scriptsize DM}}}
\def\sigmaP{\sigma_{\kern -1pt                   
                  \lower1.5pt\hbox{$\scriptstyle P$}}}
\def\sigmaT{\sigma_{\lower1.5pt\hbox{$\scriptstyle T$}}}
\def\Tsys{T_{\hbox{\scriptsize sys}}} 
\def\Vsig{V_{\hbox{\scriptsize sig}}} 

\begin{document}

%\preprint{APS/123-QED}

\title{Search for Dark Photon Dark Matter: Dark E-Field Radio Pilot Experiment}% Force line breaks with \\

\author{Benjamin Godfrey}
 
  \author{J. Anthony Tyson}%
 \email{tyson@physics.ucdavis.edu}

 \author{Seth Hillbrand}%
%\affiliation{ Authors' institution and/or address}%


\author{Daniel Polin}

\author{Jon Balajthy}
\altaffiliation[Now at ]{Sandia National Laboratories, Livermore, California, 94550, USA.}%Lines break automatically or can be forced with \\

\author{S. Mani Tripathi}

\author{Shelby Klomp}
\altaffiliation[Now at ]{Physics Department, Northwestern University, Evanston, Illinois, 60208, USA.}%Lines break automatically or can be forced with \\

\author{Joseph Levine}
%\altaffiliation[Also at ]{Physics Department, Cal State University, Chico.}%Lines break automatically or can be forced with \\

\author{Nate MacFadden}
\altaffiliation[Now at ]{NK Labs, Cambridge, Massachusetts, 02139, USA.}%Lines break automatically or can be forced with \\

\affiliation{Physics Department, UC Davis, Davis, California, 95616}

\author{Brian H. Kolner}
\author{Molly R. Smith}
\altaffiliation[Now at ]{Microsoft, Redmond, Washington, 98052.}

\affiliation{Electrical and Computer Engineering Department, UC Davis, Davis, California, 95616, USA}

\author{Paul Stucky}
\affiliation{Chemistry Department, UC Davis, 
Davis, California, 95616, USA}%

\author{Arran Phipps}
\affiliation{Physics Department, CSU East Bay, Hayward, California, 94541, USA}

\author{Peter Graham}
\author{Kent Irwin}
%\author{Arran Phipps}
\affiliation{Physics Department, Stanford University, Stanford, California, 94305, USA}


\date{\today}% It is always \today, today,

\begin{abstract}
We are building an experiment to search for dark matter in the form of dark photons in the nano- to milli-eV mass range. This experiment is the electromagnetic dual of magnetic detector dark radio experiments. It is also a frequency-time dual experiment in two ways: We search for a high-Q signal in wide-band data rather than tuning a high-$Q$ resonator, and we measure electric rather than magnetic fields. In this paper we describe a pilot experiment using room temperature electronics which demonstrates feasibility and sets useful $\epsilon\!\sim\! 10^{-12}$ over 50--300 MHz. With a factor of 2000 increase in real-time spectral coverage, and lower system noise temperature, it will soon be possible to search a wide range of masses at 100 times this sensitivity. We describe the planned experiment in two phases: Phase-I will implement a wide band, 5-million channel, real-time FFT processor over the 30–-300 MHz range with a back-end time-domain optimal filter to search for the predicted $Q\sim 10^6$ line using low-noise amplifiers. We have completed spot frequency calibrations using a biconical dipole antenna in a shielded room that extrapolate to a $5\,\sigma$ limit of $\epsilon\sim 10^{-13}$ for the coupling from the dark field, per month of integration. Phase-II will extend the search to 20 GHz using cryogenic preamplifiers and new antennas.
\end{abstract}
\maketitle


%\tableofcontents

\section{Introduction}

The physical nature of dark matter is unknown. Sensitive searches for weakly interacting massive particles (WIMPS) have found nothing ~\cite{schumann2019direct}. 
In recent years the WIMP
hypothesis has dominated searches for dark matter since a generic weak-scale thermal relic could account for all of the observed dark matter in the universe~\cite{arcadi2018waning}. Experimenters
continue to probe new WIMP parameter space by
developing larger and more sensitive detectors, however these tend to lose sensitivity when the mass of the dark matter particle is small, leaving a large range of parameter space open for exploration~\cite{2015ICRC...34....5T}. 

%We know where the dark matter is and we know how much there is. 
The 2014 P5 report~\cite{ParticlePhysicsProjectPrioritizationPanel(P5):2014pwa} emphasizes the importance of
searching for dark matter along every feasible avenue. To date, relatively little effort has been spent on detection of ultralow mass dark matter candidates where it is best described as a wave rather than a particle \cite{Battaglieri:2017aum}. This requires
development of new detectors. 

The \textit{dark photon} is a hypothetical, low-mass vector boson which has been posed as a candidate for dark matter. 
Dark photons could account for much of the dark matter, and are theoretically motivated via fluctuations of a vector field during the early inflation epoch of our universe. A relic abundance of such a particle could be produced non-relativistically in the early universe in a similar way to axions, through either the misalignment mechanism or through quantum fluctuations of the field during inflation~\cite{Arias_2012, Graham:2015rva}. 

%Further, at long wavelengths, the vector inherits the usual adiabatic, nearly scale-invariant perturbations of the inflaton,  allowing it to be a good dark matter candidate. 

In contrast to axions, a massive, inflation-produced vector boson like a dark photon would have a power spectrum that is peaked at a length scale of roughly $10^{10}$ km, and rapidly decreases in intensity at large length scales, consistent with CMB observations. Furthermore, a dark photon would adopt the adiabatic fluctuations of the inflaton making it a good dark matter candidate~\cite{Graham:2015rva}. The high phase space density required for dark photons to constitute a significant portion of the local dark matter density ($\sim$0.3 GeV/cc) 
implies that they would behave as
an oscillating field
%This dark electric field~\cite{PhysRevD.84.103501} has an amplitude of 
%$\rho_{\text{DM}}^{1/2}$ (I commented this out as we later explain it better in equation 2 -Dan)
and would oscillate with a frequency equal to the mass of the dark photon. In general, for a
theory with two U(1) symmetries, there would be some weak coupling with a corresponding term in the Lagrangian ~\cite{1986PhLB..166..196H, 1991PhLB..267..509F, Rizzo_2019}. The Lagrangian then varies from the standard
model, $\mathcal{L}_{SM}$, as shown in Eq.\ \ref{eq:lagrange}.
\begin{equation}\label{eq:lagrange}
%\begin{split}
    \mathcal{L} = -\frac{1}{4} F'_{\mu\nu} F'^{\mu\nu} + \frac{1}{2} m^2 A'_\mu A'^\mu \\
    - \frac{1}{2}\epsilon F'_{\mu\nu} F_{EM}^{\mu\nu} + \mathcal{L}_{SM}   
%\end{split}
\end{equation}

Here $m$ is the mass of the dark photon, $F_{\mu\nu}$ and $A_\mu$ are the electromagnetic field strength and gauge potential, $F'_{\mu\nu}$ and $A'_\mu$ are the dark photon field strength and gauge potential, and $\epsilon$ is the dark photon-to-electromagnetic coupling factor which must be measured. 
The mixing term between the two coupled fields is then $\frac{1}{2}\epsilon F'_{\mu\nu} F_{EM}^{\mu\nu} $.
 Through kinetic mixing, dark photons would be detectable in traditional electromagnetic searches, and $\epsilon$ can be measured.
Previous experimental bounds on $\epsilon$ from direct detection are summarized in~\cite{PhysRevD.98.030001}.
The other unknown is the mass (frequency) of the dark photon.

We are building an experiment to search for dark photons in the nano- to milli-eV mass range. This experiment is the electromagnetic dual of magnetic dark photon experiments performed by Parker \textit{et al}., ~\cite{PhysRevD.88.112004}
and Chaudhuri \textit{et al}.,~\cite{Chaudhuri_2015}. 
We have completed a feasibility test of our \textit{Dark E-field radio} experiment which has shown that the full experiment can work. Here we report on this pilot experiment. We plan to finish construction of the experiment and carry out a comprehensive search over a mass range spanning four orders of magnitude. For this, we are developing a novel time-spectrum detector which optimally detects a monochromatic signal and rejects signals varying on timescales incompatible with the model.

The remainder of the paper is organized as follows.  In Sec.~\hyperref[sec:Femtovolt]{II} we review the detection technique, and the plans for an efficient wideband real-time FFT. In Sec.~\hyperref[sec:Experiment]{III} we outline the current pilot experiment and plans for the next phase.  In Sec.~\hyperref[sec:Sensitivity]{IV} we present a sensitivity analysis describing how the limit on the kinetic coupling parameter $\epsilon$ depends on system parameters. In Sec.~\hyperref[sec:Simulation]{V} we discuss EM simulations of the response to a volume electric field in the shielded room. In Sec.~\hyperref[sec:Pilot]{VI} we describe the details of this pilot experiment and its S/N, while Sec.~\hyperref[sec:DAQ]{VII} presents our data acquisition and analysis methods. Signal injection tests of sensitivity are given in Sec.~\hyperref[sec:Injection]{VIII}. Sec.~\hyperref[sec:SignalSearch]{IX} presents the search logic and preliminary results on $\epsilon$ in this pilot experiment, and in Sec.~\hyperref[sec:Reach]{X} we outline the eventual reach of the experiment in two future phases.

\section{Femtovolt time-spectrum detector}
\label{sec:Femtovolt}
Evidence for the existence of dark matter was discovered via its gravitational effects on large scale dynamics, and new astronomical probes promise to establish additional constraints on its physical nature ~\cite{2019arXiv190201055D}. Astronomical observations \cite{Read:2014qva, 2019arXiv191014366D} jointly give the energy density of dark matter, $\rho_{\text{DM}}$, at our position in our Galaxy: 380$\pm$180 TeV/m$^3$. For dark photons in free space, converting this energy density to its corresponding observed electric field gives~\cite{Chaudhuri_2015}
\begin{equation}
\bigl|\Eobs \bigr| \approx \epsilon \sqrt{\frac{2}{\varepsilon_{0}}\rhoDM}  
\end{equation}
For our local dark matter energy density, this gives $\sim~3700$ V/m multiplied by the weak coupling factor $\epsilon$.

Because the dark field will pass through any shielding, the entire experiment is placed inside an electromagnetic shield to screen external, interfering electromagnetic sources. This shielding, however, will affect the sensitivity of the experiment. As a simple analytic example, inside a cylindrical conducting shield whose radius, $R$, is much smaller than the wavelength of the dark photon oscillations ($\lambda=h/m_{\gamma'}c$ for dark photon mass $m_{\gamma'}$), this observed field will be suppressed, giving \cite{Chaudhuri_2015}
\begin{equation}
\bigl|\Eobs \bigr| \approx \varepsilon \sqrt{\frac{2}{\varepsilon_{0}}\rhoDM}\,\left(\frac{m_{\gamma'}c}{\hbar}\right)^2R^2
\end{equation}
plus a term (that is negligible in our frequency range) $\propto m_{\gamma'} R \, v_\text{DM}$ where $v_\text{DM} \sim 10^{-3} \, c$ is the velocity of the dark matter.  In the limit where $\lambda \!\ll\! R$, the observed electric field approaches its free space value.   In the case of more complex geometries such as a shielded room with fixtures, the general trend is the same, but a numerical simulation must be performed.

As in WIMP searches, there are two unknowns: the frequency of the wave (a proxy for the mass) and its weakly coupled amplitude. We measure the induced electric field with a wideband antenna. The experiment is conducted
inside a large ($\approx$ 27.4 m$^3$) electromagnetically shielded room, searching for a weak narrowband signal between 30 MHz and 20 GHz from dark photons converting from within the shield. The antenna is polarization sensitive, enabling detection of the expected $E$-field in any direction whence aligned. The challenge is detecting a 1 ppm spectrally pure signal, varying only on 12-hour timescales
(Earth rotation), at femtovolt levels, in wideband noise. Since the frequency of the line is unknown, the search must be over as wide a spectral range as possible. For high spectral efficiency, we must also measure this small signal simultaneously at each candidate frequency. Our experiment will use a $2^{26}$-channel real time fast Fourier transform (FFT) processor using field-programmable gate arrays (FPGAs) similar to MacMahon et al.\cite{MacMahon_2018}. % except that we compute the FFT on-board.

%The former is based on cryogenic low-noise preamplifiers. The latter is based on parallel FFTs running on FPGAs yielding 62-million channel, real-time, spectrum analysis. 
In order to have high spectral efficiency plus high sensitivity, we leverage two new technologies: an ultra-low-noise radio receiving system inside a large shielded
room, plus state-of-the-art wide-band spectrum monitoring. Together with time-domain filtering, the combination will be a uniquely efficient real-time detector which can simultaneously search wide swaths of frequency for the signal from dark photons converting to detectable photons with fractional bandwidth (mass) of 1 ppm.

To test this idea, we have built a  pilot experiment inside a shielded room using a wide-bandwidth biconical antenna and room temperature preamps (Fig.~\ref{fig:room}). To date, we have
performed two proofs of concept: First, we detect a 150 pV/m converted dark photon proxy signal injected into the shielded room using an rf signal generator (outside the shielded room) connected to a low-gain dipole inside the room over the course of a 1-month integration.

Second, we do a two-month long run between 50–
300 MHz with a $Q$ of $10^5$ looking for single bin anomalies on top of the rf background. However, both runs suffer from ultra-low spectral efficiency because of the limited frequency span of the commercial real-time spectrum analyzer currently used in our pilot experiment. To cover a wide spectral range we must integrate over each narrow span and then step to the next span.

For the next phase of the experiment, our planned time-spectrum detector with cryogenic preamps will solve that problem by addressing the twin issues of simultaneous wide spectrum sampling and time invariant signal filtering, enabling a sensitive search for dark photons over a large range of effective mass.

\section{The Experiment}
\label{sec:Experiment}
We propose to develop a novel 100\% spectral efficiency detector utilizing multiple special antennas, with cryogenic preamps, and construct three efficient FFT spectral monitoring systems~\cite{2016JAI.....541001H}, along with the software for time-signature filter and data analysis. A time-frequency signal-matched filter significantly enhances the signal-to-
background ratio by demanding that the signal occupy only one spectral bin and be constant in time for hours. Because a monochromatic constant signal transmits no information, no intentional signal has this property.

Instrumenting two separated shielded rooms we plan to run simultaneous searches in multiple frequency bands, integrating for a month, covering a vast frequency range, ultimately THz. Spectral data will be stored frequently so
that any candidate detection can be analyzed for time dependence because the power spectrum of a true dark matter $E$-field signal will not vary on short timescales.

\begin{figure}[ht]
  \centering
  \includegraphics[width=0.48\textwidth]{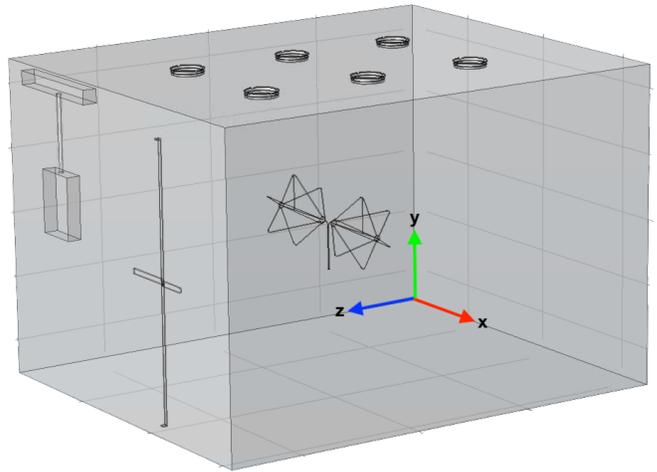}
  %\fbox{\rule[-.5cm]{4cm}{4cm} \rule[-.5cm]{4cm}{0cm}}
  %\vspace{-4ex}
  \caption{The Dark E-field radio experiment inside a large electromagnetic shielded room, searching for a narrowband signal between 30 MHz and 20 GHz from dark photons converting inside the shield. The antenna is placed in the center of the room. Features that affect the modes of the room are shown including light fixtures, electrical box, and door latch. All of these impact the antenna’s response to wall currents. The output from the antenna is fed into a low noise amplifier inside the shield, whose buffered output is connected to a wideband real-time spectrum analyzer for data processing.}
  \label{fig:room}
\end{figure}

Figure \ref{fig:room} shows the current 30--300 MHz setup in a $3.05 \times 2.45 \times 3.67$ m commercial shielded room with over 100 dB isolation. For the pilot experiment all conducting features in the room which could measurably affect our simulations over a 50--300 MHz range have been included in the solid model. We have done EM simulations versus
frequency of the induced currents in the walls due to the weakly converting dark photon field and validated the predicted antenna voltage by injected signal tests. At frequencies above 200 MHz the room suppression of the signal is small, and the antenna transfer function (antenna factor) approaches that of free-space.

% \vspace{-2ex}
%~\ref{fig:limits}.  
%The merged averaging scheme will allow access to frequency ranges over three
%decades with less than -10dB non-linearity in response. 

%A 100\% efficient parallel FFT search for a $Q\!\sim\!10^6$ line simultaneously over 1 GHz of spectrum, with the unique time signature, will be possible.
Before describing the results of our pilot experiment, in this section we describe plans for the next phases of the experiment. In Phase-I (post-pilot), a wide band FFT over
the 30--300 MHz region, with a back end optimal filter, will be used for the initial search for the predicted $Q\!\sim\!10^6$ line, using low-noise amplifiers.
%This experiment is now running. 
 Phase-II will search up to
20 GHz and will use cryogenic preamps and a novel antenna design. We can ultimately cover a factor of 10,000 in mass range, at levels of sensitivity up to 10$^5$ times better than current astrophysical limits. 
%With the addition of a permanent magnet in the shielded room we would be able to simultaneously search for axions~\cite{wagner2010search, Crisosto:2019fcj, Ouellet:2018beu}.

\section{Sensitivity analysis}
\label{sec:Sensitivity}
Of course the direction of a monochromatic $E$-field from dark photons is unknown. Although we do not record phase, the receiver is sensitive to $E$-field direction due to the polarization sensitivity of the antenna. For example a dipole has peak sensitivity to an $E$-field aligned with its axis. We mount the antenna such that it is most sensitive to the east-west component of an $E$-field, so that if a signal is detected, its amplitude will be modulated on 12-hour periods by the Earth’s rotation.

The sensitivity of the receiver in this pilot experiment is limited by the thermal noise of the low-noise preamplifier, which dominates the thermal emission from the wall. The voltage presented at the input to the preamp by thermal
radio emission from the walls is reduced by two factors relative to the noise voltage generated in the preamp: the emissivity of the wall, and the antenna factor (see Sec.~\ref{sec:Simulation}).

The measurable quantities are the total power, $P_{T}$, and the system noise power, $\Psys$. The noise power can also be measured by averaging the measured power in nearby frequency bins. The signal power, $\Psig,$ adds linearly with the noise power and thus can be found by subtracting the noise power from the total power
\begin{equation}
   \Psig =  P_{T}- \Psys= \Vsig^2  \,
    \hbox{Re} \bigl\lbrace Z \bigr\rbrace / |Z|^2
\end{equation}
where $\Vsig$ is the measured RMS voltage and $Z$ is the impedance of the antenna. 
We use a balun to match the impedance of the antenna to the transmission line and therefore
$ \hbox{Re} \bigl\lbrace Z \bigr\rbrace / |Z|^2 \approx 1/|Z|$.
The signal voltage is related to the electric field at the position of the antenna, $\vec{E}_x$ by the \textit{antenna factor} \textit{AF}
\begin{equation}
    AF \equiv \biggl| \frac{\vec{E}_x}{\Vsig} \biggr|
\end{equation}
This response, \textit{AF}, is defined as the electric field component coupling to the antenna divided by the corresponding voltage developed at the antenna terminals. \textit{AF} has units of $\textrm{meters}^{-1}$.
% PRD review: re-inserted following sentence
For example, for an antenna with 50 Ohm termination $$ AF = \frac {9.73} {\lambda G^{\frac{1}{2}}} $$ where $\lambda$ is the wavelength in meters and G is the antenna numerical gain over isotropic. 

In free space, the dark matter energy density is related to the measured electric field via

\begin{equation}
\rhoDM = \frac{\varepsilon_{0}}{2\epsilon^{2}}\left|\vec{E'}\right|^{2} 
\end{equation}
where $\epsilon$ is the small kinetic mixing parameter between the dark photon and electromagnetism and $\varepsilon_{0}$ is the permittivity of free space. 
%defined by
%\begin{equation}
%\varepsilon_{0} = \frac{e^{2}}{4\pi\alpha\hbar c}
%\end{equation}
%\brian{I don't think it is necessary to define 
%$\epsilon_0$ in this paper}
%
From this, the signal power is related to the local dark matter energy density according to
%\begin{equation}\label{eq:PowerDef}
%\Psig=\frac{\epsilon^2}{\left(AF\right)^2|Z|\varepsilon_{0}} \rhoDM
%\end{equation}

\begin{equation}\label{eq:PowerDef}
  \Psig=\frac{2\epsilon^{2}}{\varepsilon_{0}\left(AF\right)^2 |Z|}\rhoDM  
\end{equation}

The uncertainty in a noise measurement on a narrow band signal, $\sigmaT$, is given by the Dicke radiometer equation
 \cite{Dicke:1946}

\begin{equation}
    \sigmaT \approx \frac{\Tsys}{\sqrt{\DeltanuRF\tau}}
\end{equation}
where $\Tsys$ is the system noise temperature. The bandwidth, $\DeltanuRF$, times the integration time, $\tau$, gives the number of trials in the integration. The random uncertainty in the measured power, $\sigmaP$, is therefore given by
\begin{equation}
    \sigmaP \approx \kB \DeltanuRF\sigmaT
\end{equation}
This uncertainty applies to both the measurement of total power and to the measurement of the baseline power. The total power is measured for a single bin, whereas the baseline power is measured using a large number of bins, so the statistical uncertainty in the signal power is approximately given by
\begin{equation}
    \sigmaP \approx \kB \sigmaT\approx \frac{\kB \Tsys}{\sqrt{\DeltanuRF\tau}}
\end{equation}
\begin{figure}[ht]
  \centering
  \includegraphics[width=0.48\textwidth]{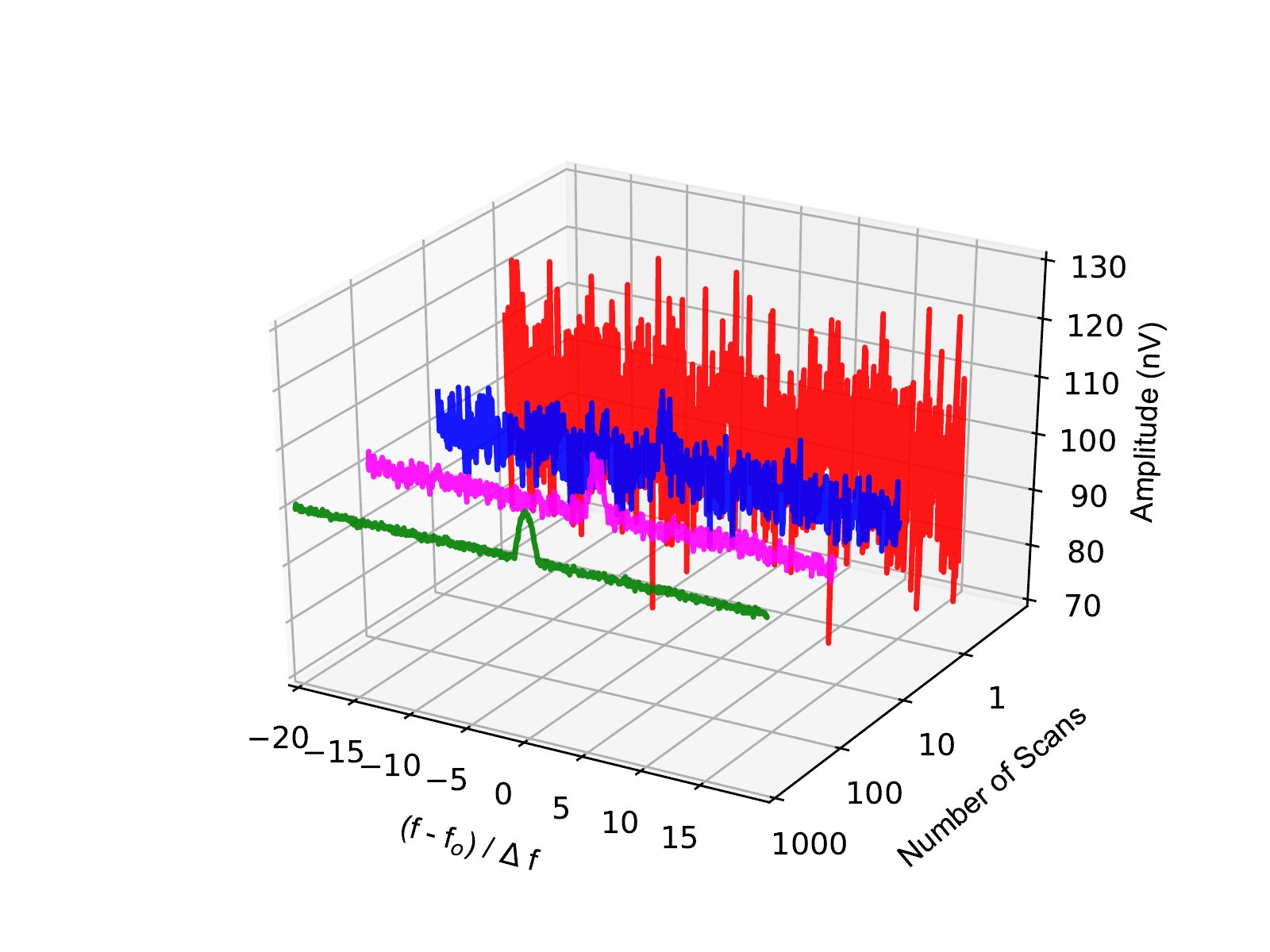}
  %\fbox{\rule[-.5cm]{4cm}{4cm} \rule[-.5cm]{4cm}{0cm}}
  %\vspace{-4ex}
  \caption{SNR dependence on number of scans averaged for a 10nV signal, relative to a 100 nV long-time averaged noise floor, at a frequency of f$_0$. The red, blue, magenta, and green curves represent the signal after 1, 10, 100, and 1,000 scans, respectively. The width of this signal is assumed to be $\Delta f= f_0/10^6$. Note that even though the average noise level goes down like the square root of time [Eq. (\ref{eq:SNR})], the limit on the scalar coupling constant, $\epsilon$, goes as the quarter root of time [Eq. (\ref{eq:LOD})].}
  \label{fig:fakesignal}
\end{figure}

\begin{figure}[ht]
  \centering
  \includegraphics[trim=+0.7cm 0 0 0, width=0.48\textwidth]{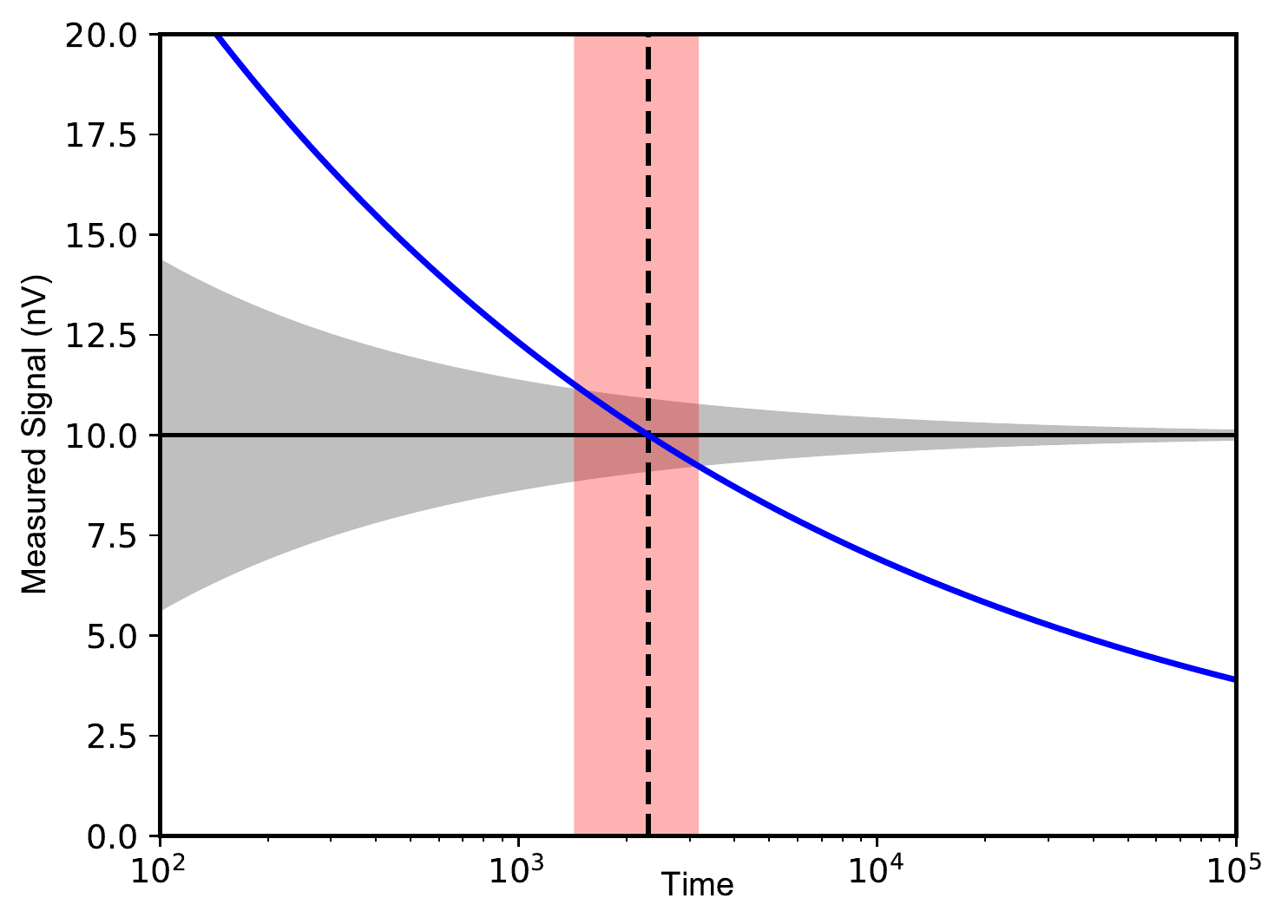}
  %\fbox{\rule[-.5cm]{4cm}{4cm} \rule[-.5cm]{4cm}{0cm}}
  %\vspace{-4ex}
  \caption{Time dependence of a detection. The gray shaded region shows the predicted $\pm$1$\sigma$ band of the integrated voltage measurement from a 10 nV signal as a function of time. The solid blue line shows the 3$\sigma$ exclusion limit as a function of time. The black dashed line and shaded region shows the predicted time until a 3$\sigma$ detection.}
  \label{fig:limitsetting}
\end{figure}

Figures~\ref{fig:fakesignal}~and~\ref{fig:limitsetting} illustrate the predicted behavior of the averaged spectral data and eventual measurement error in the event of a detection of a 10~nV dark matter signal. Figure~\ref{fig:fakesignal} shows the time-evolution of the baseline noise near a detection. As the baseline averages down, the exclusion limit will decrease as (number of scans)$^{1/4}$. As it approaches the 10~nV signal, the additional power from the dark photon field will manifest as a worsened exclusion limit around the central frequency, $f_0$. Once the 5$\sigma$ threshold is crossed, the uncertainty on the measurement will decrease as the inverse square root of time.

Figure~\ref{fig:limitsetting} depicts the integration time dependence of the 5$\sigma$ exclusion limit, and of the predicted integrated voltage measurement from a 10~nV dark matter signal. The limit curve again decreases like the quarter root of the number of scans, and a discovery will occur when the measurement rises above this line. The exclusion line enters the $\pm$1$\sigma$ band of the voltage measurement at 1,428 scans and exits at 3,272 scans. This means that although the expected time until detection of such a signal is 2304 scans, the actual time will vary by roughly 40\%. 

The uncertainty in the coupling constant due to system noise, $\sigma_{\epsilon}$, can be found using standard error propagation
\begin{equation}\label{eq:SNR}
    \begin{split}
        \sigma_{\epsilon}^2 &= \left[\frac{\partial}{\partial \Psig}\left(\sqrt{\frac{\left(AF\right)^2|Z|\varepsilon_{0} \Psig}{2 \rhoDM}}\right)\right]^2\sigmaP^2\\[2ex]
        \sigma_{\epsilon} &= \kB \Tsys\sqrt{ \frac{\left (AF \right)^{2}|Z| \varepsilon_{0}\DeltanuRF}{8\rhoDM \Psig\tau}}
    \end{split}
\end{equation}
If a positive detection is defined as a signal equal to a constant multiple of the system noise, $\xi \in \mathbb{R}^{+}$, the limit of detection for a given integration time is given by
\begin{equation}\label{eq:LOD}
\begin{split}
    \frac{\epsilon}{\sigma_{\epsilon}} &= \frac{2 \Psig}{\kB \Tsys} \sqrt{\frac{\tau}{\DeltanuRF}}\equiv \xi \\[2ex]
    \epsilon \Bigl|_{\hbox{\scriptsize SNR}=\xi} &= \left(\frac{\Delta \nu}{\tau} \right)^{1/4} \sqrt{\frac{\xi \kB \Tsys\left( AF \right)^{2} |Z| \varepsilon_{0}}{2 \rhoDM}}
\end{split}
\end{equation}

Equation (\ref{eq:SNR}) describes how the uncertainty of a given dark photon signal decreases with integration time, while Eq. (\ref{eq:LOD}) [obtained by substituting Eq. (\ref{eq:PowerDef})] describes how the limit~of~detection is improved with integration time. The former scales like the inverse square root of time, and can be thought of as an expression of the central limit theorem, while the latter scales as the inverse quarter root of time. 
For example, in order to reduce the limit of detection measured using an hour of data by a factor of 10, the integration time would have to be increased to about 1 year.

From Eq. (\ref{eq:LOD}), a factor of four decrease in system temperature is
equivalent to a factor of 16 longer integration per span. Clearly, cryogenic preamplifiers are necessary, and the limit on $\epsilon$ will improve until the preamplifier noise temperature becomes small relative to the effective temperature of the thermal radiation from the walls.
As discussed in Sec. \ref{Pilot}, there is a further trade-off between integration time per FFT span and the number of FFT spans.

\section{EM simulation of response}
\label{sec:Simulation}
\begin{figure}[ht]
  \centering
  \includegraphics[width=0.48\textwidth]{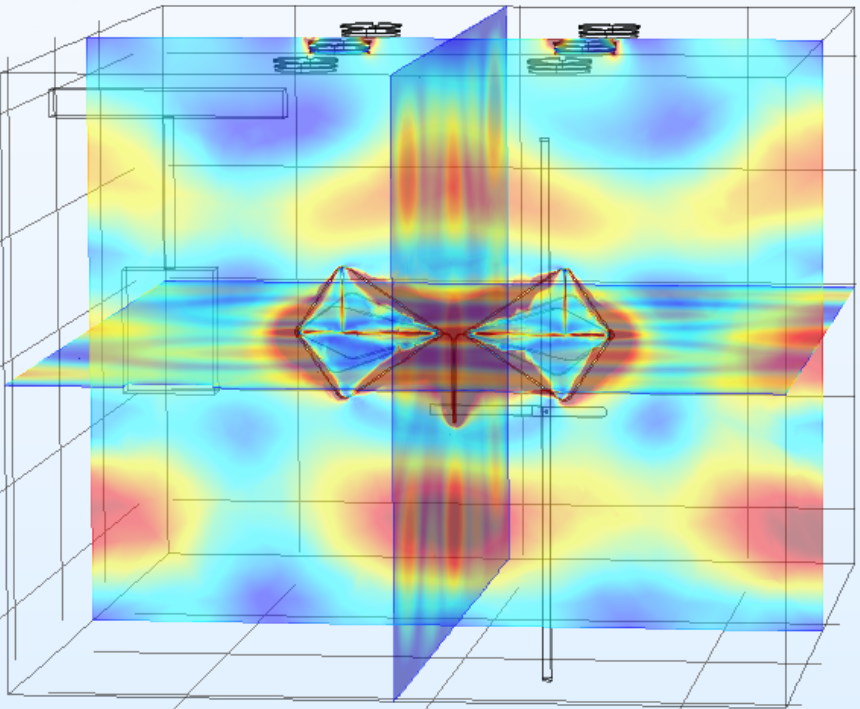}
  %\fbox{\rule[-.5cm]{4cm}{4cm} \rule[-.5cm]{4cm}{0cm}}
  %\vspace{-4ex}
  \caption{COMSOL  simulation  of  the TE035 room mode at 275.9 MHz. The interior of the room with  our  biconical  antenna  is shown filled with heat-map slices of the E-field strength.  Dark blue corresponds to zero field and dark red to $10^3$  V/m}
  \label{fig:mode}
\end{figure}

The antenna and the shielded room are sufficiently complex that an analytic derivation of the effective antenna factor is impractical. As a cross validation, we calculated the antenna factor analytically for a thick dipole in a rectangular waveguide and in free space, and confirmed the results using 
COMSOL electromagnetic simulation software.
Using both COMSOL and CST EM software (\cite{COMSOLRef}, \cite{CSTRef}) we carried out simulations of the response of our antenna to a converted volume $E$-field in the shielded room. As mentioned above, this response is the antenna factor ($AF$). For our antenna in a shielded room, the boundary conditions of the conducting walls mean that the $AF$ exceeds the free-space value (antenna response suppressed) at frequencies for which the wavelength is large compared to the size of the shielded room. At frequencies above this cutoff, the $AF$ is quite unlike the free space $AF$ because of the strong coupling of the antenna to the modes of the shielded room. In fact, the antenna-room system becomes essentially a larger EM detector -- providing gain (lower $AF$) about a factor of 10 in excess of the isolated dipole over wide frequency ranges~\cite{hill2009}.  
The model for the shielded room and antenna is taken from measurements with precision of 0.2\% of the shortest wavelength.
An example of the mode structure in the shielded room from a COMSOL simulation is shown in Fig.~\ref{fig:mode}.

%{\sl COMSOL} and {\sl CST} differ in the methods by which one can specify a source field. {\sl COMSOL} allows a volume $E$-field source, while {\sl CST} does not, requiring the extra step of simulating a current in the walls which then induces an $E$-field in the shielded room. 
% PRD review: re-inserted following 2 sentences
At most frequencies the strong coupling with room modes effectively raises the detection volume to a significant fraction of the volume of the room. Small antenna factor corresponds to high sensitivity.

\begin{figure}[ht]
  \centering
  \includegraphics[width=0.48\textwidth]{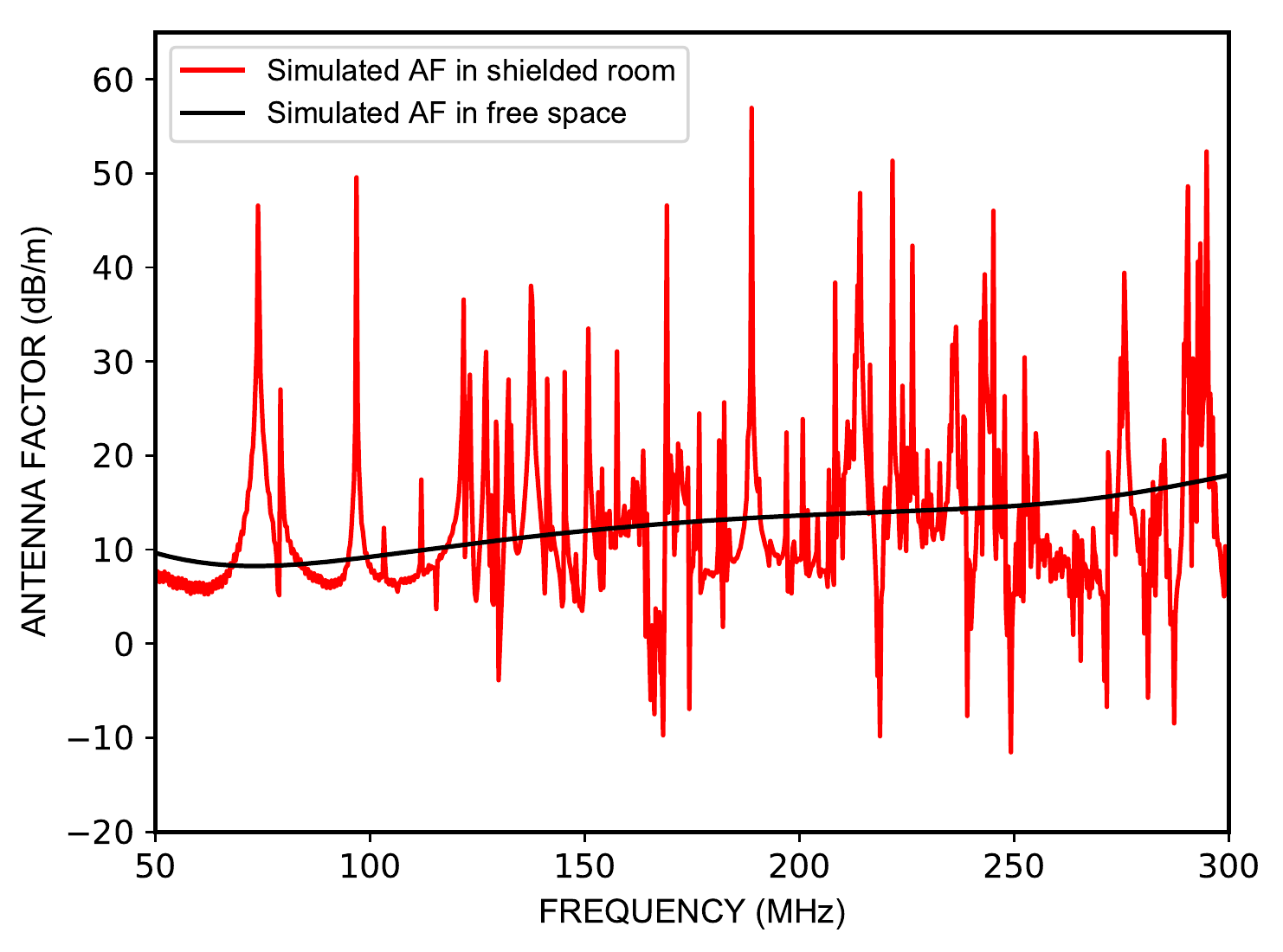}
  %\fbox{\rule[-.5cm]{4cm}{4cm} \rule[-.5cm]{4cm}{0cm}}
  %\vspace{-4ex}
  \caption{CST simulation of the antenna factor of the biconical antenna in the shielded room as a function of frequency with a broadband 50:180$\Omega$ balun matching network (red). Also shown is the simulation of the free-space $AF$ (black) emphasizing how the room and antenna become a strongly coupled system.}
  \label{fig:antfac}
\end{figure}

An EM simulation of the antenna factor for our biconical dipole antenna in the shielded room, using CST, is shown  in Fig.~\ref{fig:antfac}. This simplified antenna plus room model ``empty room'' allows comparison to analytically derived modes.  Even the presence of the antenna in the room causes mode-mode coupling.  
%For example, the wide dip at 75 MHz is several coupled modes in the 70-80 MHz region.  The peak at 140 MHz is caused by a mode that puts an E field null at the antenna.
%\vspace{-1ex}
%~\ref{fig:limits}.  

\section{Pilot Experiment}\label{Pilot}
\label{sec:Pilot}
The pilot study provides a proof of concept of experimental design. Data are collected from 50-300 MHz using the setup outlined in Fig.~\ref{fig:room}. For this feasibility study, a biconical antenna, low noise, room temperature amplifiers, and a commercial real-time spectrum analyzer (RTSA: Rigol Model RSA5065-TG) are used. 

In the pilot experiment, memory limitations of the current RTSA require splitting up the entire spectral range into a series of sequential spans, which reduces acquisition efficiency. 
This span is, in part, defined for a given window type, by a span per resolution ratio, \textit{SRR}. Given a required resolution, \textit{Q}, (defined as frequency over resolution bandwidth), a center frequency, \textit{CF}, the span of a scan can then be calculated from
\begin{equation}
 \textrm{Span} = CF\left[\frac{Q}{SRR} + \frac{1}{2}\right]^{-1}
 \end{equation}
 
As an example, for a fixed resolution of $10^{5}$ and a 1024-point Kaiser window, this requires 454 separate scans to cover the entire range from 50-300 MHz. This loss in efficiency will be mitigated in the ultimate real-time data acquisition system with 100\% efficiency,
%described in Section~\ref{subsec:ADC} that will 
allowing simultaneous acquisition across the entire spectral range.  

%\vspace{-1ex}
%~\ref{fig:limits}.  
%\begin{figure}[h]
%  \centering
%  \includegraphics[width=0.4\textwidth]{tex/CurrentPlots/antenna.jpg}
  %\fbox{\rule[-.5cm]{4cm}{4cm} \rule[-.5cm]{4cm}{0cm}}
  %\vspace{-4ex}
%  \caption{The biconical dipole antenna suspended in the middle of the shielded room. This antenna has a relatively flat free space spectral response between 30 and 300 MHz.}
%  \label{fig:antenna}
%\end{figure}

    In order to make external EM interference subdominant to contributions from thermal and amplifier noise, the shielded room must have greater than 100 dB shielding across the entire spectral range surveyed. Experiments were done to verify that the shielded room met these requirements by transmitting a signal of known frequency and constant power from the outer lab room and observing the antenna response when the door to the room was open versus closed. Care was taken not to overdrive amplifiers, and calibrated attenuators were necessary. These data confirm that the dominant contribution to the noise floor is not from externally generated EM sources. In addition, the outer lab itself is shielded. Nevertheless, a few high power known signals are detected inside the shielded room. Thus, monitoring of the spectrum outside the shielded room is a necessary feature of our experiment. Any signal detected in the shielded room must be at least 100 dB stronger outside if it is due to external rf.
    
%\begin{figure}[ht]
%  \centering
%  \includegraphics[width=0.5\textwidth]{GivenVal%uesAndSpotCheck_Corrected.pdf}
%  \caption{Electric field attenuation of shielded room from 100 kHz to 100 GHz. The red dots show spot tests done at 110 and 912 MHz to verify that performance of the shielded room is as specified by the manufacturer.}
%  \label{fig:shielding}
%\end{figure}

\begin{figure}[ht]
  \centering
  \includegraphics[trim=+0.5cm 0 0 0, width=0.48\textwidth]{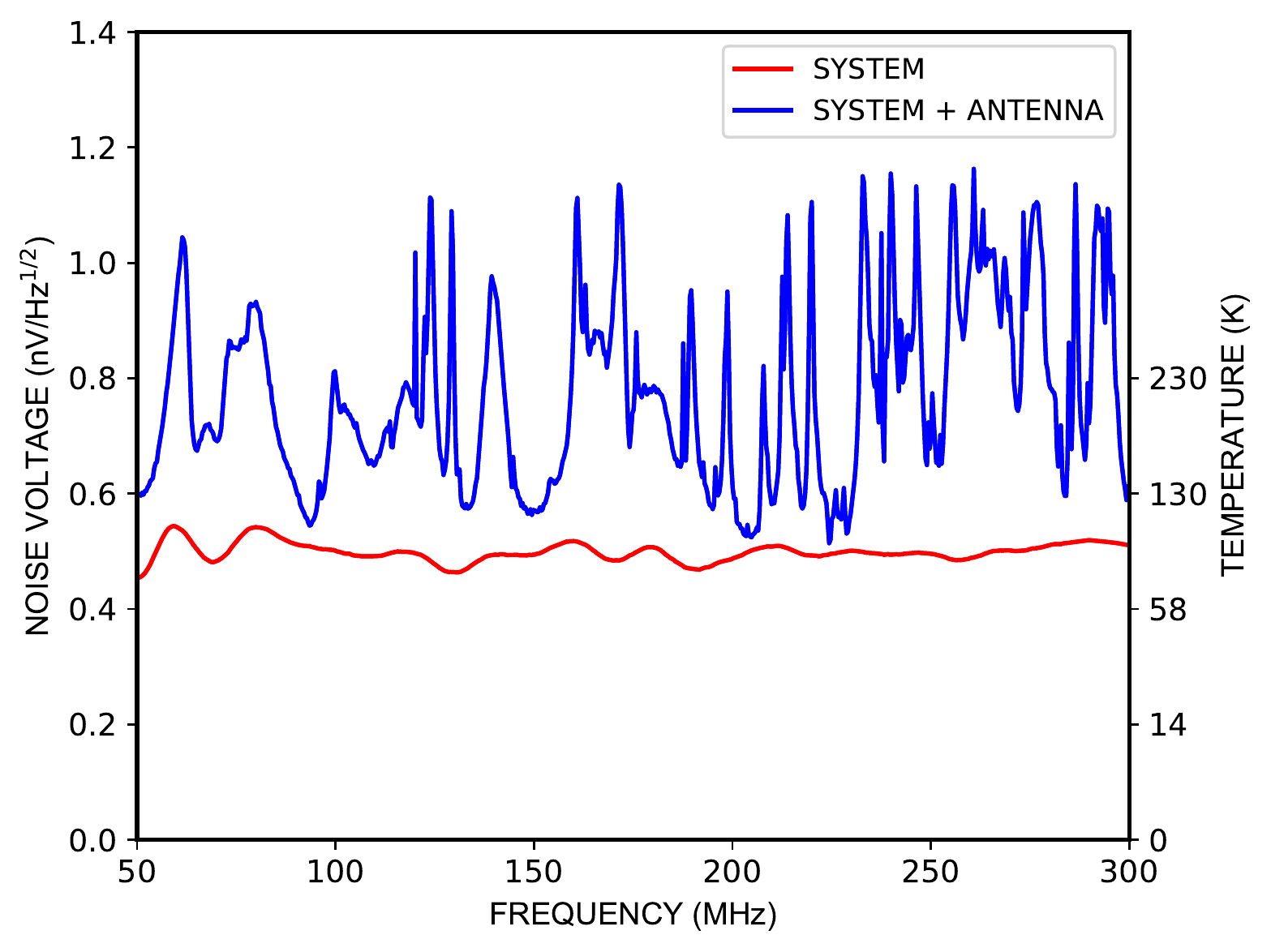}
  %\fbox{\rule[-.5cm]{4cm}{4cm} \rule[-.5cm]{4cm}{0cm}}
  %\vspace{-4ex}
  \caption{ Blue: The spectrum of noise in the shielded room referred to the input of the preamp, showing modes coupled to the polarization of the bicon antenna. %These modes are labeled as TE$_{\textrm{xyz}}$ where \textit{xyz} designate the directions shown in Figure~\ref{fig:room}. 
  This shielded room thermal noise is on top of the wideband white noise of the room temperature preamplifier. 
  Red: System noise temperature referenced to 50$\Omega$ (the input impedance of the preamp) for short-circuited preamp input. Future runs will use cooled preamplifiers. }
  \label{fig:antenna_noise}
\end{figure}

The rf noise in the shielded room is non-zero and originates from two sources. First, the 20\% emissivity of the 290 K walls creates a thermal background which tends to pile up at the room resonances: $\frac{1}{2} \kB T$ per degree of freedom. This is the strongest noise source in the case of a low noise preamp.
%This contributes xxx per mode after accounting for the antenna factor. 
The second weaker contribution comes
from the noise coupling out of the input port of the preamp itself. Broadband noise from the input port of the preamp can couple to the antenna and radiate into the shielded room exciting room modes. Both of these room noise signals are in addition to the wideband noise of the preamp. This induced nonwhite room noise then couples to the antenna into the preamp, creating a $\sim$0.5 nV/Hz$^{1/2}$ added component (modulo the coupling of the mode to the antenna polarization) on top of the 100 K noise temperature preamp white noise spectrum of 260 pV/Hz$^{1/2}$. A noise spectrum covering the full range of the Phase-I bicon antenna is shown in Fig.~\ref{fig:antenna_noise}. The measured RMS noise-voltage spectral density vs frequency is shown, referred to the output of the antenna. Also shown for reference is the short-circuit noise spectral density of the preamp.
  
\section{data acquisition and analysis}\label{sec:DAQ}

The goal of the experiment is to search for a \textit{Q} $\approx 10^{6}$, time invariant signal on less than 12-hour time scales, that is completely submerged in noise. This is done using the following process.
For each span, small subsets of data are bin-averaged together. Any spans that have signals 5$\sigma$ above the baseline are thrown out. This works as a data filter to remove high-amplitude transient signals, reducing unwanted noise.

To search for a monochromatic signal (one spectral bin) in the presence of noise, the amplitude of each bin must be understood relative to the noise floor. This noise floor is not flat, due to a combination of shielded room and instrument effects. To resolve this, each bin-averaged span is passed through an optimized high-pass filter to remove low frequency baseline shape %(sixth-order high pass Butterworth with an 80-bin cutoff) 
to remove low frequency baseline shape.

A post-FFT filter for the background noise floor is used that assigns weights to frequency bins near the trial frequency. This helps to determine whether a measured signal is significantly above the noise floor. The shape of this filter is optimized using signal injection tests by finding the window shape that maximizes the final measured signal-to-noise ratio for nearby background noise.

Memory limitations in the RTSA used in this pilot experiment require dividing the 50--300 MHz range into a series of smaller scans called spans, each of which has the required fractional spectral resolution. These narrow spectral spans are further subdivided in time into smaller, 10-second time intervals to maximize the ability to veto spurious external noise while also being written sufficiently infrequently to avoid write-to-disk bottlenecks leading to losses in scanning efficiency. All data files are written out in a plain text format and saved to disk. Afterwards, they are converted to fixed-size, binary format (HDF5) files for data processing. The resulting time-frequency database of time-tagged sequential spectral spans enables a series of detection validation tests. Any detection of a monochromatic signal may be examined for its time dependence and compared in amplitude with the expected 100 dB stronger signal on a spectrum analyzer and broadband antenna outside the shielded room.

\section{signal injection tests}
\label{sec:Injection}
\begin{figure}[ht]
  \centering
  \includegraphics[width=0.48\textwidth]{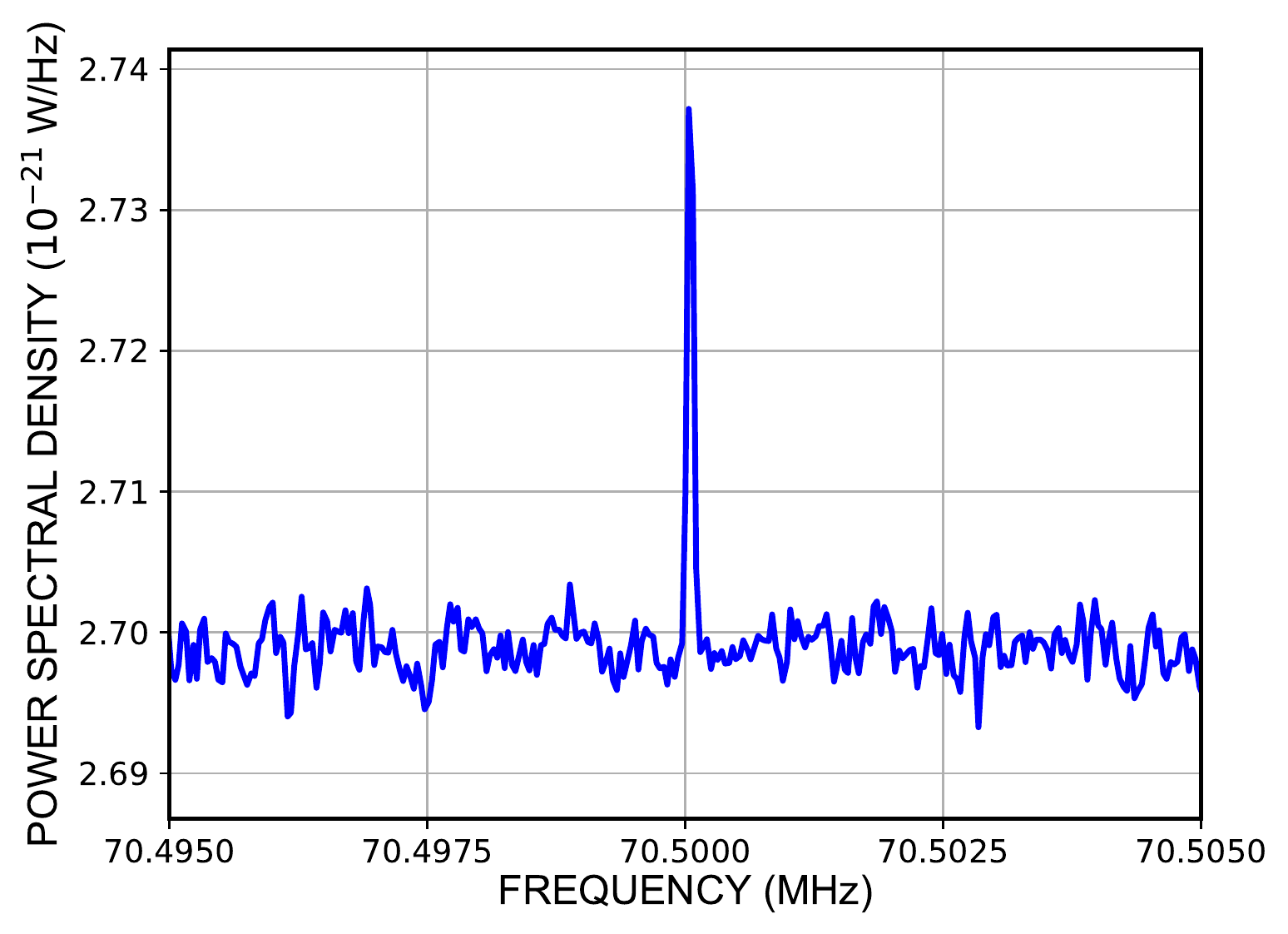}
  %\fbox{\rule[-.5cm]{4cm}{4cm} \rule[-.5cm]{4cm}{0cm}}
  %\vspace{-4ex}

  \caption{Power spectrum showing a 22$\sigma$ detection at 70.5 MHz after $10^6$ seconds integration (frequency span=10 kHz). Total signal power received by the antenna is 3.5 $\times 10^{-23}$W, which is a factor of 730 below our detection threshold without averaging. This signal injection test demonstrates the sensitivity of our pilot system.}
  \label{fig:inject}
\end{figure}

As a test of the sensitivity of the Phase-I system, we inject a small signal at one frequency and integrate the FFT over a narrow bandwidth. Figure \ref{fig:inject} shows an example of the resulting spectrum.
For this test, we inject a small signal at 70.5 MHz into the shielded room using a small bow-tie antenna. This signal is well below the noise floor of the receiver. A narrow-band sweep from 70.48-70.51 MHz was then performed many times to extract the signal from the noise. Since the span is narrow, we can scan in real time, performing a nearly 100\% efficient
FFT over this frequency range. 
The result for $10^6$ sec integration shows an amplitude of 42~pV (RMS) referred to the input of the preamplifier. 
Using the standard deviation of the baseline in the narrow-band sweep, and the known antenna factor, a 5$\sigma$-limit on the kinetic mixing parameter, $\epsilon$, of $6.4 \times 10^{-13}$ between 70.49 and 
70.51 MHz can be inferred. 
%Over the course of a month, using the current setup,  this limit 
%could be reduced to  $1.3 \times 10^{-13}$.
 
%\vspace{-1ex}
%~\ref{fig:limits}.  

%\subsection{Measurement of limiting $\epsilon$  vs integration time}

 We test the limiting detectable signal vs integration time and find that the peak detectable power scales inversely as the square root of integration time. %(far from the quantum limit). 
 As discussed in Section~\ref{sec:Sensitivity}, the limiting $\epsilon$ then scales as the fourth root of integration time. A study of the average of many scans with a weak injected signal
 %is shown in Figure~\ref{fig:SNRvsTime}, where the cumulative power SNR is plotted vs the integration time.  
 demonstrated that the expected behavior from Eq. (~\ref{eq:SNR}) is observed.

%\begin{figure}[h]
%  \centering
%  \includegraphics[width=0.49\textwidth]{tex/CurrentPlots/SNRvsTime_10-12-20.pdf}
  %\fbox{\rule[-.5cm]{4cm}{4cm} \rule[-.5cm]{4cm}{0cm}}
  %\vspace{-4ex}
%  \caption{SNR versus time of a 70.5MHz, $3.5 \times 10^{-23}$W signal injected into shielded room. Each data point shows the SNR after a number of hours of averaging. A fit to this data then shows how SNR grows as a function of time. This should go as the square root of integration time as per the Dicke radiometer formula. 
  %However, gain variations on long time scales reduce this slightly.
%  }
%  %$t^{0.5}$. } 
%  \label{fig:SNRvsTime}
%\end{figure}

There is a trade off between channel width (needed to attain high $Q$), number of spans, and integration time in the search phase vs followup phase for any detected signal.  In
the search phase, detection of a monochromatic signal buried in noise is the goal. This puts emphasis on integration time per Hz. However, because the frequency of the signal is unknown, the maximum range must be covered during the search phase. For example halving the FFT resolution doubles the spectral efficiency (half the number of required spans) for a given run time. Splitting the spans during data acquisition into sequential time samples enables additional filtering for the expected constant signal.

Once candidate signals are detected above threshold, a run at high efficiency may be made with much higher spectral resolution in spans centered on the candidate frequencies. A surviving candidate may then be validated by demanding a non-detection in the spectrum monitoring system with an antenna outside the shielded room.

\section{Search for a signal}
\label{sec:SignalSearch}
In this section we outline the process of searching for a narrowband constant signal, and describe the results of our pilot Phase-I search. Due to the low efficiency of searching 50-300 MHz we consider two types of search: 
\begin{enumerate}
    \item A 100\% efficient search confined to several narrow spans at several spot frequencies
    \item A broadband search covering 50-300 MHz but at $\sim {1}/{400}$ the efficiency of the spot limits.
\end{enumerate} 

\begin{figure}[ht]
    \centering
    \includegraphics[width=0.47\textwidth]{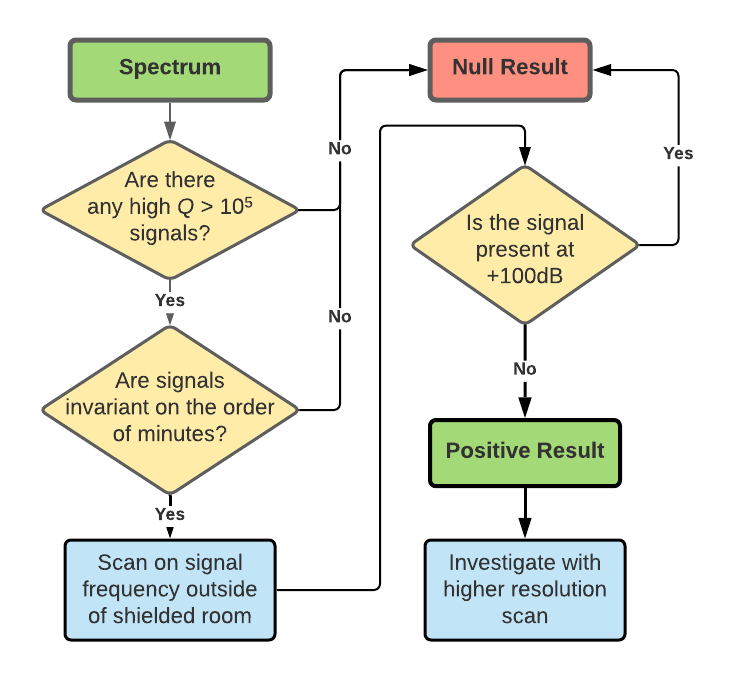}
    \caption{Flow chart describing the logic of the search process. Starting with the data from the measured spectrum, the figure shows analysis of a detected signal, and the automated methodology for eliminating false positive signals.}
    \label{fig:signalflowchart}
\end{figure}
After an initial run is performed, and our data are processed, we must determine where any potential high $Q$ signals that match our criteria originate. The methodology for this process is outlined in Fig.~\ref{fig:signalflowchart}. 

The rf spectrum outside the shielded room is dominated by electromagnetic signals which contain information. Thus, in general, they do not exhibit extremely high $Q$ and constant amplitude.  The nature of any signal detected in the shielded room may be revealed by comparing the signal measured inside the shielded room to the signal outside the room. External electromagnetic signals in our frequency range should be attenuated inside the shielded room by 100 dB or more. 

This search logic flowchart applies to our Phase-I and Phase-II experiments as well.  Once a candidate high-$Q$ apparently constant signal is detected there are a series of investigations that will occur.  Using a higher resolution smaller span targeted run the shape of the line could be investigated, and any time-variation detected. Based on those results, as well as non-detection outside the shielded room, the next step will be to confirm the signal in an identical setup some distance away. We are preparing that infrastructure. Finally, the next obvious step will be to focus
on that frequency with the vastly superior sensitivity of a superconducting quantum limited detector of the kind employed by the Stanford group~\cite{2019arXiv190608814P}.

\begin{figure}[ht]
\centering
\includegraphics[trim=+0.7cm 0 0 0, width=0.5\textwidth]{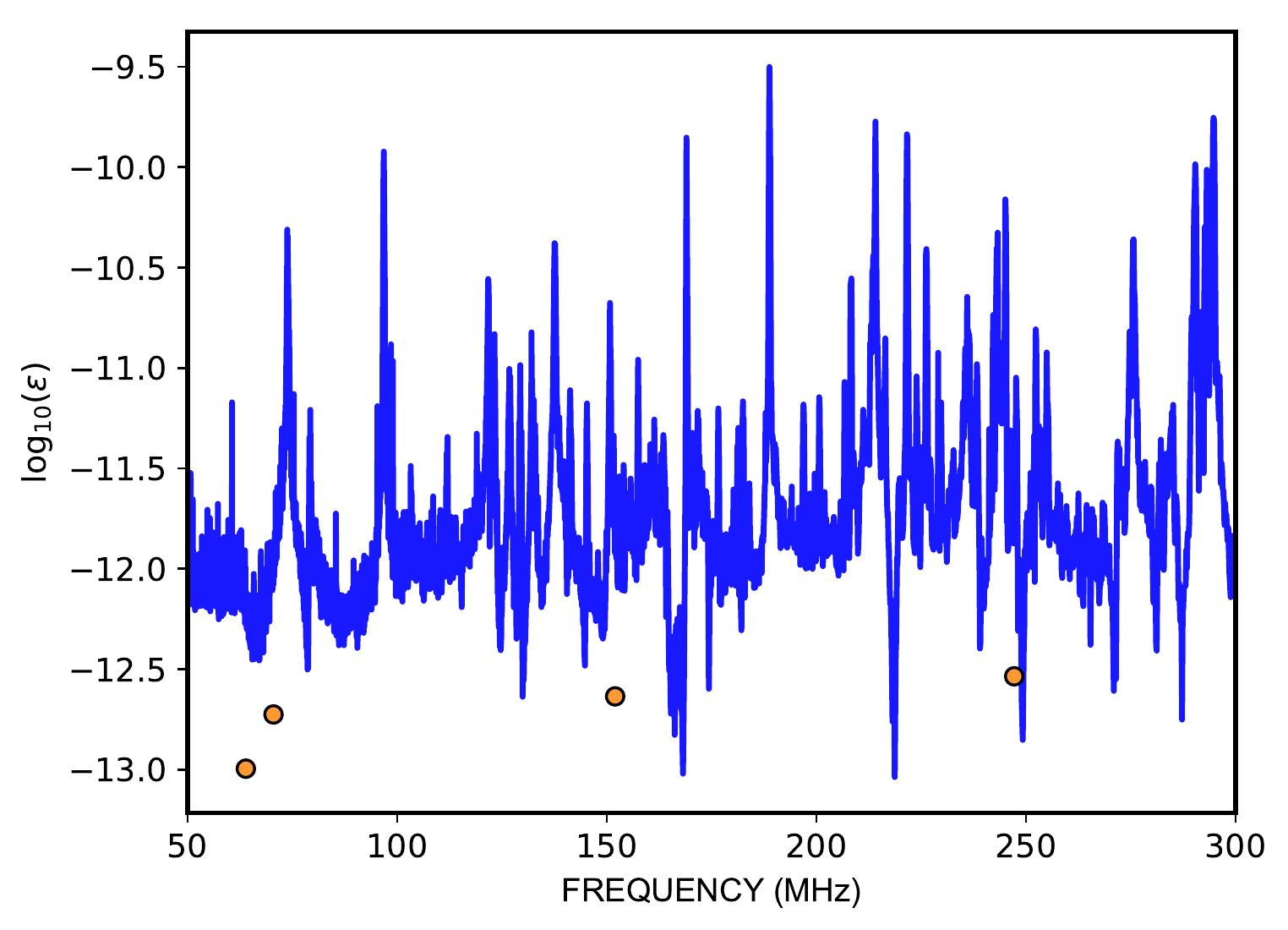}
\caption{Reach of the pilot experiment after short test integrations. The blue curve is for only 3.8 hours of real-time data collection.  The orange dots show spot measurements at 63.9, 70.5, 152, and 247 MHz taken with at least $5\times 10^{5}$sec of real-time data extrapolated to 1 month of data acquisition (assuming the current noisy receiver), and are offset from the blue curve by the expected amount due to the different integration times.}
\label{fig:EpsilonLimits}
\end{figure}

\subsection*{Limits on \texorpdfstring{$\mathbf{\epsilon}$}{epsilon} 50--300 MHz}
The first run of the pilot experiment was done over the course of two months scanning between 50 and 300 MHz. The goal was a proof of concept.
%of experimental principle and design.
Figure~\ref{fig:EpsilonLimits} shows the results of approximately 3.8 hours of real-time data collection from 50-300 MHz using the system show in Fig.~\ref{fig:room}. This short effective integration resulted from the dead time of the RTSA inherent in the sequential frequency scans. The data taking procedure is outlined in Sec.~\ref{sec:DAQ}. Conversion from voltage to electric field was done using the modeled antenna factor shown in Fig.~\ref{fig:antfac}. This was converted to an $\epsilon$ limit by looking at the standard deviation of the electric field at each bin and calculating the 5$\sigma$ limit. 

\begin{figure}[ht]
\centering
\includegraphics[trim=+0.5cm 0 0 0, width=0.49\textwidth]{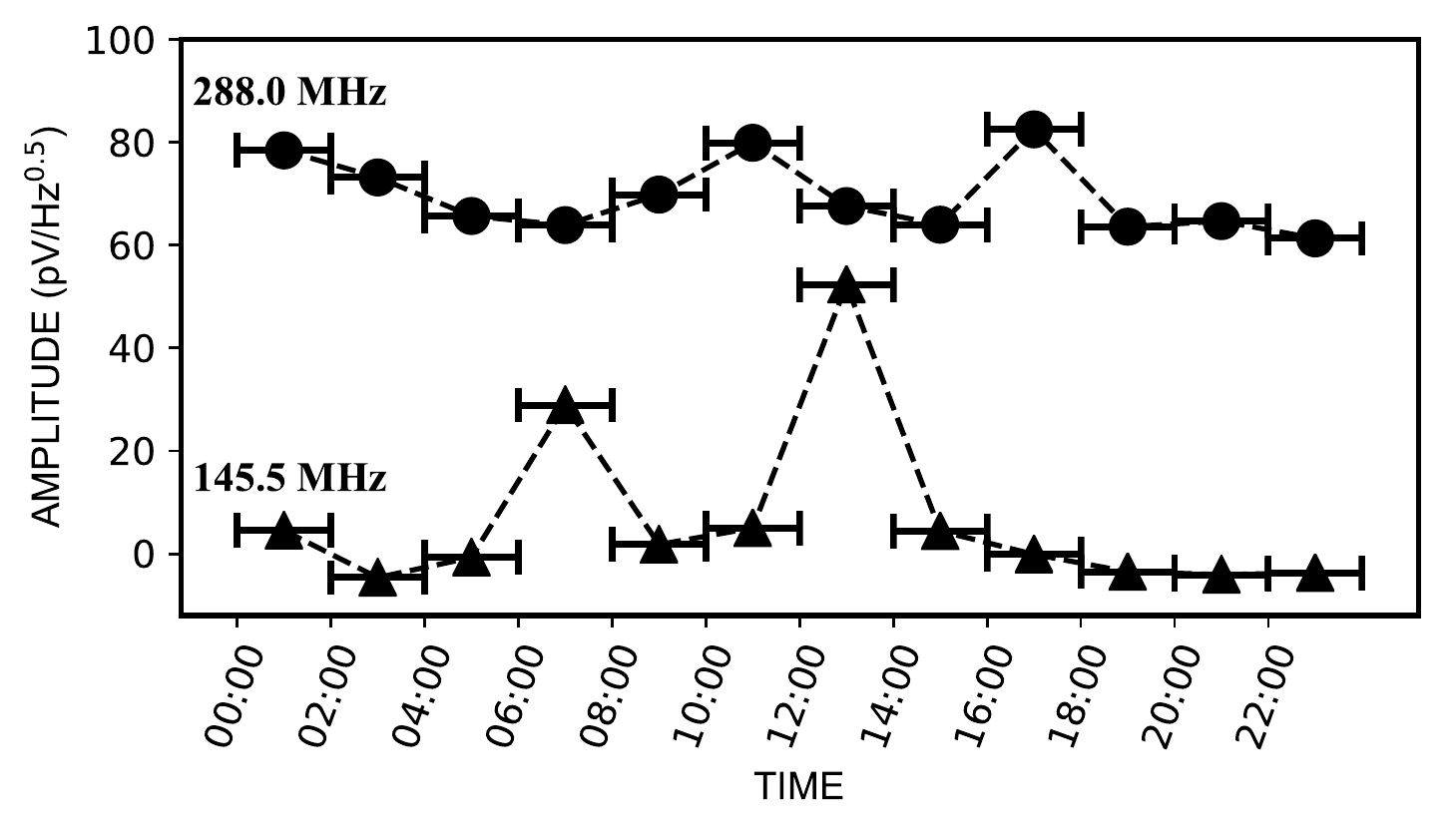}
\caption{Voltage above baseline of $\sim$19 minutes of data over a 2-hour time window. 
  At 145.5 MHz, a weekly ham radio net on the 100W K6JRB repeater at 12:30 p.m. is clearly detected. This signal can be eliminated due to short term time-variance. At 288.0 MHz, a possible candidate of unknown origin is shown. This is eliminated by checking the signal outside the shielded room, finding it 110 dB stronger. %(Figure~\ref{fig:ExcludingSignal}).}
}
\label{fig:CandidateSignals}
\end{figure}

This search produced approximately 130 bins where there was a signal of interest (equal to 0.03 \% of all bins in the search). Many of these signals are quickly excluded via Step 2 in Fig.~\ref{fig:signalflowchart}. Signals that remain after this process are generally excluded by Step 3. However, care must be taken
because some of these candidates are transient. Looking at the dependence of the voltage noise above baseline on the time of day the signal gives a further handle on whether to exclude these as possible candidates. As described in Fig.~\ref{fig:signalflowchart},  splitting up the data into short time bins, enables monitoring signals versus time in order to discriminate transient interference from candidate dark photons. Figure~\ref{fig:CandidateSignals} highlights the transient nature of one of these candidate peaks at 145.5 MHz, which is the location of a weekly ham radio net. 
%A real dark photon candidate would be mostly independent of the time of day that the data was taken so minimal variation of the observed SNR should be observed.

\begin{figure}[ht]
\centering
\includegraphics[trim=+0.5cm 0 0 0,width=0.49\textwidth]{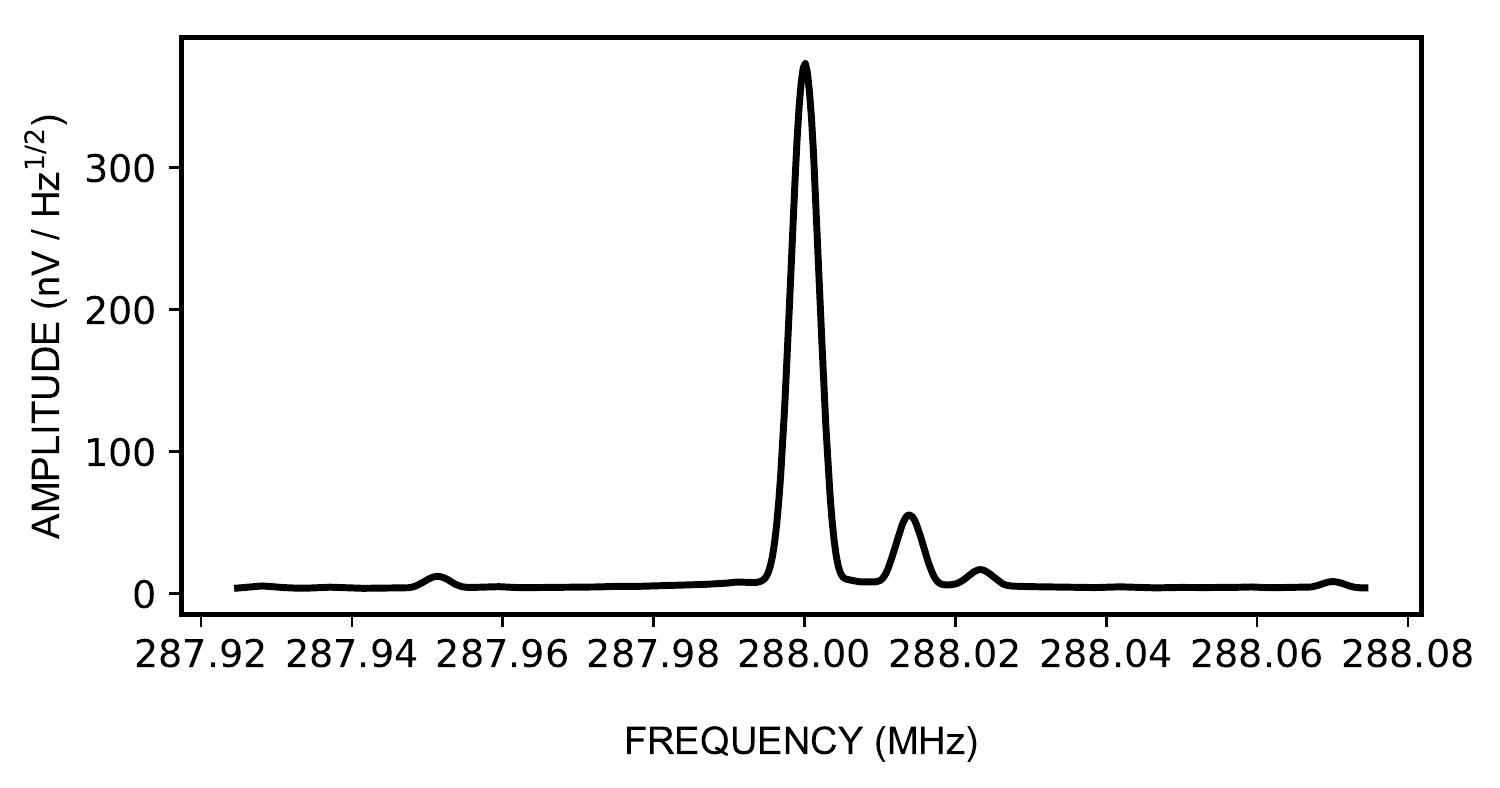}
\caption{A scan around a candidate signal at 288 MHz done with a wide-bandwidth Vivaldi antenna outside the shielded room showing 1,000 seconds of real-time data. The signal is 100 dB stronger than our shielded room detection. This excludes this signal as a candidate. Other EM peaks, also from the same antenna, are evident but are below the sensitivity of the experiment when the shielded room door is closed.}
\label{fig:ExcludingSignal}
\end{figure}

The other candidate detection at 288 MHz was eliminated by comparing with the rf spectrum outside the shielded room. While there is variation in the SNR at 288 MHz, the candidate is only conclusively excluded by doing a narrow-band sweep around the candidate frequency outside the shielded room. The result of this is shown in Fig.~\ref{fig:ExcludingSignal}. From this, it is clear that this signal has terrestrial origins.

\subsection*{Limits on \texorpdfstring{$\mathbf{\epsilon}$}{epsilon} at four spot frequencies}\label{subsec:SpotLimits}
Four narrow-span scans around 63.9, 70.5, 152, and 247 MHz were chosen as a second calibration of sensitivity as shown by the orange dots in Fig.~\ref{fig:zoomlimits}.
For each spot check, at least half a million seconds of data were obtained and used to compute a 5$\sigma$ limit on $\epsilon$. This limit is extrapolated to a month using Eq. (\ref{eq:LOD}). The locations of spot frequencies were chosen to sample across the entire 50--300 MHz span as well as to investigate the sensitivity of the experiment in regions with/without room modes. Span widths were selected to ensure a resolution of at least one part in 10$^{5}$ and are the same as used in the pilot experiment (Fig.~\ref{fig:EpsilonLimits}).
% PRD review: re-inserted following sentence
It is expected that near room modes (see Fig.~\ref{fig:antenna_noise}), sensitivity will improve. However, this is complicated by the properties of the antenna itself (see Figure~\ref{fig:antfac}).  The resulting antenna factor is effectively a convolution of the two and can only be calculated via EM simulation.

\section{Reach of the proposed experiment}
\label{sec:Reach}
\begin{figure}[ht]
  \centering
  \includegraphics[trim=+1cm 0 0 0, width=0.465\textwidth]{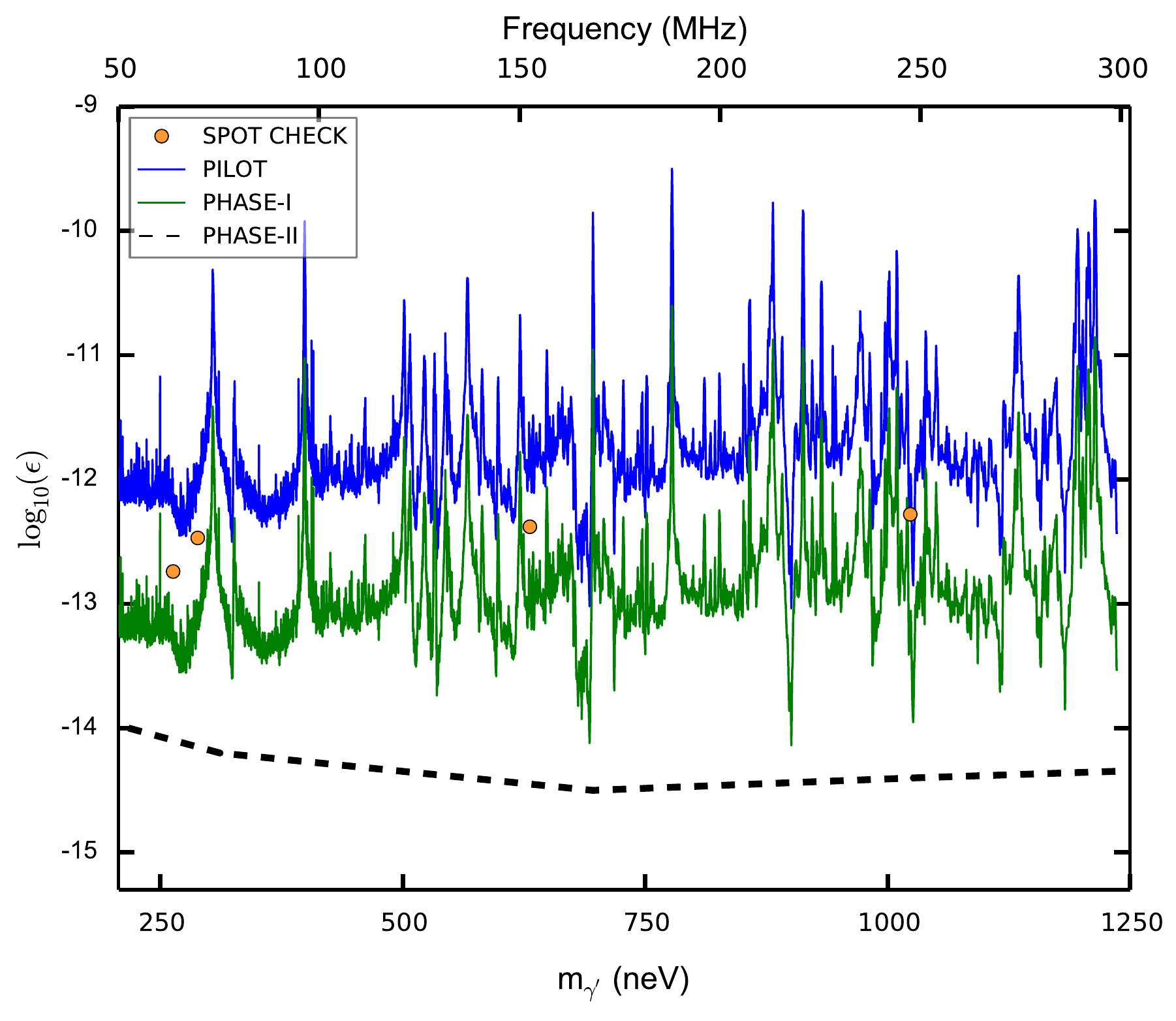}
  %\fbox{\rule[-.5cm]{4cm}{4cm} \rule[-.5cm]{4cm}{0cm}}
  \vspace{-4ex}
  \caption{Limit plot showing the reach of the experiment from 50--300 MHz in several stages. Pilot (blue) shows 5$\sigma$ limits after 3.8 hours of data collection using the proof-of-concept setup. Phase-I (green) shows 5$\sigma$ extrapolated limits using current antenna with LNA noise temperature and RBW improved by a factor of 2 and 10, respectively, after 1-month of real-time data acquisition. Spot checks (orange) show measurements around 63.9, 70.5, 152, and 247 MHz taken with at least 5 $\times 10^{5}$sec of real-time data in the pilot experiment. Unlike Fig.~\ref{fig:EpsilonLimits}, these values are not extrapolated to one-month. As expected, these limits lie between the measured and projected limits from the pilot and Phase-I runs, respectively. Phase-II (black) shows the projected 5$\sigma$ limits for a month-long data run using a cryogenic amplifier.}
  \label{fig:zoomlimits}
\end{figure}

\begin{figure}[ht]
  \centering
  \includegraphics[trim=+1cm 0 0 0, width=0.465\textwidth]{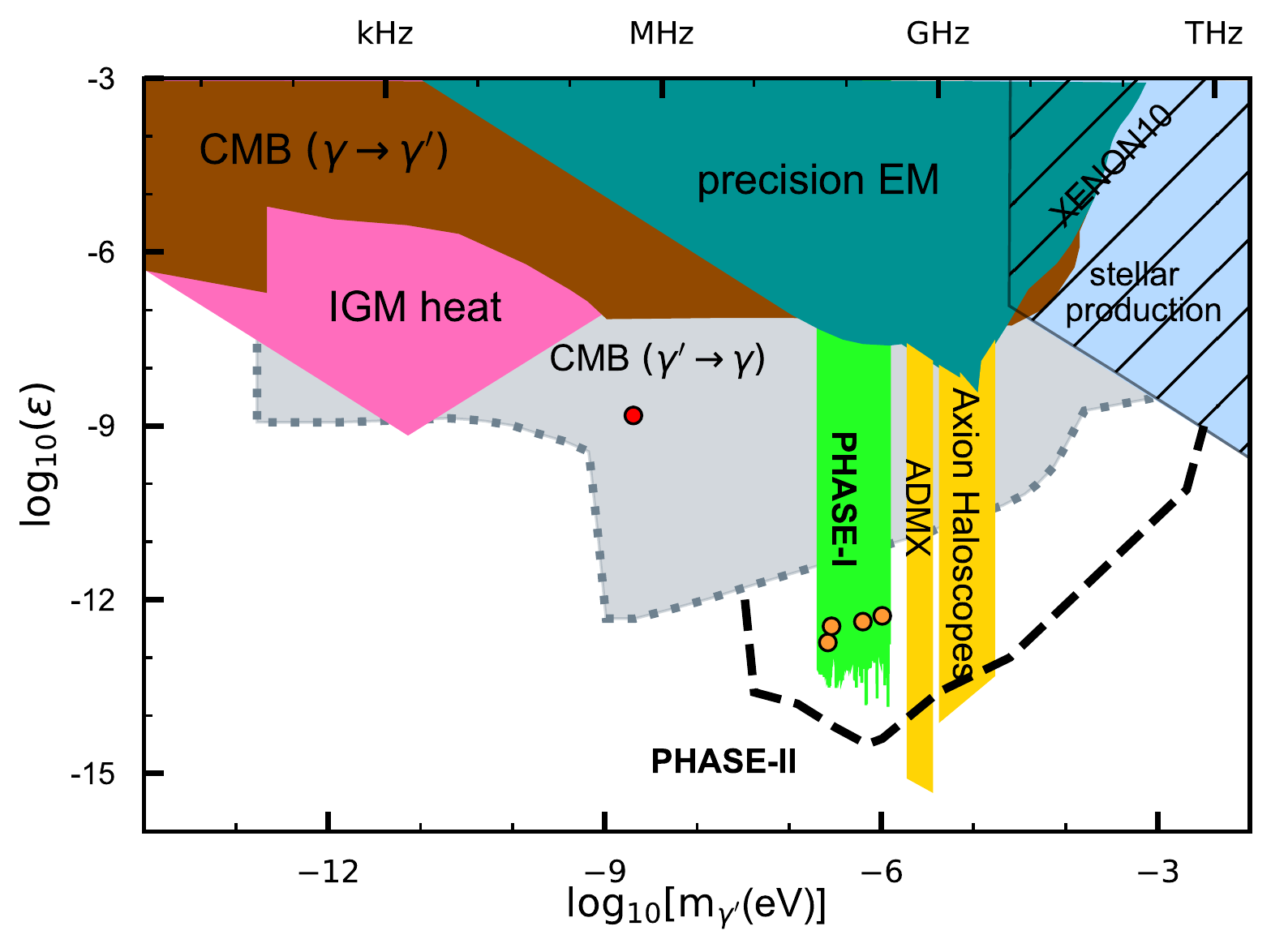}
  %\fbox{\rule[-.5cm]{4cm}{4cm} \rule[-.5cm]{4cm}{0cm}}
  \vspace{-4ex}
  \caption{Projected reach of the Dark E-Field Radio Experiment in the next phase. The weak kinetic mixing factor is plotted vs dark photon mass in eV. Regions excluded by astrophysics are shown. The light ${\gamma^\prime}-{\gamma}$ CMB region is a model-dependent constraint above which hidden photons would not account for the total dark matter density.
For reference, planned ADMX axion searches are shown in yellow, because those experiments may detect dark photons \cite{wagner2010search, Arias_2012}. Recent results, \cite{ADMX2018, ADMX2020}, may improve these limits. The orange points in Phase-I show calibrated exclusion regions at 4 spot frequencies 
at 5~$\sigma$ measured in the current noisy pilot experiment. The red dot shows the point exclusion limit measured by Phipps \textit{et al}. (2019). Our Phase-I and -II limits are based on 1 month integration. Phase-I green region shows 5$\sigma$ extrapolated limits using current antenna with improved LNA. Phase-II are cryogenic experiments covering wider ranges.}
  \label{fig:limits}
\end{figure}

Using data from our pilot experiment, we can estimate the constraining power vs integration time in various frequency (dark photon mass) bands. This is crosschecked with 100\% efficiency narrow-span scans around 63.9, 70.5, 152, and 247 MHz as a second calibration of sensitivity. The limit plot in Fig.~\ref{fig:zoomlimits} shows measured and projected limits between 50--300 MHz. Figure~\ref{fig:limits} overlays these projections across a much wider parameter space. Recently, using a superconducting resonator and SQUID detector Phipps \textit{et al}.~\cite{2019arXiv190608814P} have obtained a limit point at 0.49 MHz, shown by the red dot.
%We use our signal injection data to calibrate the sensitivity:  6 fV per month integration, at 5$\sigma$.  
 
Some discussion of the astrophysical limits and their robustness is in order. 
The $\gamma$--$\gamma'$ CMB bound (brown) comes from conversion of normal CMB photons to dark photons and does not require that the dark photons be dark matter or even have a cosmic abundance.
The $\gamma'$--$\gamma$ bound (dark photon to regular photon) is model dependent because it requires that the dark photon abundance exist at the time of conversion~\cite{Arias_2012}.  
Not only is it assumed that the dark photon
is dark matter today, it assumes that it also existed back in the early universe, in most cases before matter-radiation equality when we do not actually have any evidence that dark matter existed. For the dark photon to be dark matter today but not exist back then implies that it would have to be produced sometime between then and now (actually between then and matter-radiation equality). Thus the $\gamma'$--$\gamma$ bound is weaker than other astrophysical bounds, and speculative. The IGM heat bound arises from dark photons converting in the ionized intergalactic medium, below the
plasma frequency and heating it above observed IR bounds~\cite{2015JCAP...12..054D}.  

This pilot experiment uses room temperature preamplifiers and low efficiency spectrum acquisition. Since the error on $\epsilon$ scales like the system temperature, as well as the square root of the real-time spectral coverage, the next obvious step is to transition to cooled preamplifiers and the GHz RTSA described above. Improvements in the antennas are also planned. The dashed black line represents our forecast for the second phase of this experiment. It relies on a series of technical improvements in the different frequency bands, ranging from cryogenic preamplifiers at GHz frequencies to low temperature bolometers at THz, and a range of frequency dependent antenna designs. For example, the next antenna after the Phase-1 biconical dipole is a Vivaldi antenna with gain covering 300 MHz to 5 GHz. Ultimately, this experiment will be limited by the warm walls of the shielded room. As with any broadband search, once a signal is detected and validated with multiple detectors, the next step would be to focus on that frequency with more sensitive quantum-limited high-$Q$ detectors.
%At higher frequencies where the antenna is smaller, a series of antennas in small cooled shields may offer a path forward. At lower frequencies, where the wavelength exceeds the size of the shielded room, a larger room brings much larger increases in sensitivity than a lower system temperature (equation~\ref{eq:LOD}).  

%\newpage

\section*{Acknowledgments}
\label{sec:Acknoledgments}
We thank the Nokia and Xilinx for major donations. This project is supported by the Brinson Foundation and DOE grant DE-SC0009999.
NSF grants PHY-1560482 and PHY-1852581 supported three REU students (N.M., S.K., and J.L.). J.B. and B.G. were partially supported by the Nuclear Science and Security Consortium, funded by the DOE under the grant DE-NA00003180.
We acknowledge discussions with Barbara Neuhauser and Werner Graf. %who are setting up remote detectors for the next phase of this experiment. 
J.A.T. thanks Saptarshi Chaudhuri, Markus Luty, Jeremy Mardon, Surjeet Rajendran and Greg Wright for helpful discussions.  Some of this work was done at Aspen Center for Physics, which is supported by National Science Foundation grant PHY-1607611. We thank John Conway and Eric Prebys for reviewing an early version of this paper, and an anonymous reviewer for their helpful comments.

\newpage
\bibliography{darkradio}% Produces the bibliography via BibTeX.

\end{document}